\documentclass[conference]{IEEEtran}
\IEEEoverridecommandlockouts

\usepackage{cite}
\usepackage{amsmath,amssymb,amsfonts}
\usepackage{algorithmic}
\usepackage{graphicx}
\usepackage{textcomp}
\usepackage{xcolor}
\usepackage{siunitx}

\usepackage{threeparttable}
\usepackage{booktabs}

\usepackage{acronym}
\usepackage{float}
\usepackage{multirow}
\usepackage{soul}
\usepackage{comment}
\usepackage{makecell} %for multi-row cells

\usepackage{xcolor}

\usepackage{array}
\newcolumntype{P}[1]{>{\centering\arraybackslash}p{#1}}
\newcolumntype{M}[1]{>{\centering\arraybackslash}m{#1}}

\def\BibTeX{{\rm B\kern-.05em{\sc i\kern-.025em b}\kern-.08em
    T\kern-.1667em\lower.7ex\hbox{E}\kern-.125emX}}
 \DeclareRobustCommand*{\IEEEauthorrefmark}[1]{%
   \raisebox{0pt}[0pt][0pt]{\textsuperscript{\footnotesize #1}}%
 }
  \usepackage{fancyhdr}
  \fancypagestyle{mahmood}{%
   \fancyhf{} % clear all fields
   
   \fancyhead[C]{}
  }%

  \makeatletter
  \let\ps@IEEEtitlepagestyle\ps@mahmood
  \makeatother
\makeatletter
\def\ps@IEEEtitlepagestyle{%
  \def\@oddfoot{\mycopyrightnotice}%
  \def\@oddhead{\hbox{}\@IEEEheaderstyle\leftmark\hfil\thepage}\relax
  \def\@evenhead{\@IEEEheaderstyle\thepage\hfil\leftmark\hbox{}}\relax
  \def\@evenfoot{}%
}
\def\mycopyrightnotice{%
  \begin{minipage}{\textwidth}
  \centering \scriptsize
  \copyright 2023 IEEE.  Personal use of this material is permitted.  Permission from IEEE must be obtained for all other uses, in any current or future media, including reprinting/republishing this material for advertising or promotional purposes, creating new collective works, for resale or redistribution to servers or lists, or reuse of any copyrighted component of this work in other works.
  \end{minipage}
}

\begin{document}
\acrodef{HR}{Heart Rate}
\acrodef{BLE}{Bluetooth Low Energy}
\acrodef{DSP}{Digital Signal Processing}
\acrodef{NN}{Neural Network}
\acrodef{EEG}{Electroencephalography}
\acrodef{FFC}{Flexible Flat Cable}
\acrodef{ULP}{Ultra Low Power}
\acrodef{IMU}{Inertial Measurement Unit}
\acrodef{FR}{frame rate}
\acrodef{PMIC}{Power Management Integrated Circuit}
\acrodef{WULPUS}{Wearable Ultra Low-Power Ultrasound}
\acrodef{US}{Ultrasound}
\acrodef{ULP}{ultra low-power}
\acrodef{AFE}{analog front-end}
\acrodef{RF}{radio frequency}
\acrodef{ML}{machine Learning}
\acrodef{FPGA}{Field-Programmable Gate Array}
\acrodef{LVDS}{low-voltage differential signalling}
\acrodef{SPI}{serial peripheral interface}
\acrodef{MCU}{microcontroller unit}
\acrodef{ADC}{analog to digital Converter}
\acrodef{HDL}{hardware description language}
\acrodef{DMA}{direct memory access}
\acrodef{HMI}{Human-Machine Interface}
\acrodef{HD}{hardware design}
\acrodef{AD}{algorithm development}
\acrodef{FPS}{frames per second}
\acrodef{DAS}{delay-and-sum}
\acrodef{TGC}{time-gain compensation}
\acrodef{FP}{floating point}
\acrodef{BW}{bandwidth}

\acrodef{NE16}{Neural Engine 16}
\acrodef{CNN}{convolutional neural network}
\acrodef{RNN}{recurrent neural network}
\acrodef{FFT}{Fast Fourier Transform}
\acrodef{SSVEP}{Steady State Visually Evoked Potential}
\acrodef{BCI}{Brain-Computer Interface}
\acrodef{PULP}{Parallel Ultra Low Power}
\acrodef{SoC}{System on Chip}
\acrodef{BLE}{Bluetooth Low Energy}
\acrodef{EEG}{electroencephalography}
\acrodef{EMG}{electromyography}
\acrodef{ECG}{electrocardiogram}
\acrodef{PPG}{photoplethysmogram}

% Commands for having comments in colors.

\title{BioGAP: a 10-Core FP-capable Ultra-Low Power IoT Processor, with Medical-Grade AFE and BLE Connectivity for Wearable Biosignal Processing
\thanks{The authors acknowledge support from the Swiss National Science Foundation (Project PEDESITE) under grant agreement 193813}
}

\author{
  \IEEEauthorblockN{
    Sebastian Frey\IEEEauthorrefmark{1},
    Marco Guermandi\IEEEauthorrefmark{2}\IEEEauthorrefmark{4},
    Simone Benatti\IEEEauthorrefmark{2}\IEEEauthorrefmark{3},
    Victor Kartsch\IEEEauthorrefmark{1},
    Andrea Cossettini\IEEEauthorrefmark{1},
    Luca Benini \IEEEauthorrefmark{1}\IEEEauthorrefmark{2}\\
    }
    
    \vspace{0.2cm}

  \IEEEauthorblockA{\IEEEauthorrefmark{1}Integrated Systems Laboratory, ETH Z{\"u}rich, Z{\"u}rich, Switzerland}
  \IEEEauthorblockA{\IEEEauthorrefmark{2}DEI, University of Bologna, Bologna, Italy}
  \IEEEauthorblockA{\IEEEauthorrefmark{3}DIEF, University of Modena and Reggio Emilia, Reggio Emilia, Italy}
  \IEEEauthorblockA{\IEEEauthorrefmark{4}Greenwaves Technologies, Grenoble, France}
}

\IEEEoverridecommandlockouts
\IEEEpubid{\makebox[\columnwidth]{979-8-3503-4647-3/23/\$31.00~\copyright2023 IEEE\hfill} \hspace{\columnsep}\makebox[\columnwidth]{ }}

\maketitle
\IEEEpubidadjcol

\begin{abstract}
Wearable biosignal processing applications are driving significant progress toward miniaturized, energy-efficient Internet-of-Things solutions for both clinical and consumer applications.
However, scaling toward high-density multi-channel front-ends is only feasible by performing data processing and \ac{ML} near-sensor through energy-efficient edge processing.
To tackle these challenges, we introduce BioGAP, a novel, compact, modular, and lightweight (6g) medical-grade biosignal acquisition and processing platform powered by GAP9, a ten-core ultra-low-power SoC designed for efficient multi-precision (from FP to aggressively quantized integer) processing, as required for advanced ML and DSP. BioGAP's form factor is 16x21x14~mm$^3$ and comprises two stacked PCBs: a baseboard integrating the GAP9 SoC, a wireless \ac{BLE} capable SoC, a power management circuit, and an accelerometer; and a shield including an \ac{AFE} for ExG acquisition. Finally, the system also includes a flexibly placeable \ac{PPG} PCB with a size of 9x7x3~mm$^3$ and a rechargeable battery ($\phi$ 12x5~mm$^2$). We demonstrate BioGAP on a \ac{SSVEP}-based \ac{BCI} application.
We achieve 3.6~$\mu J/sample$ in streaming and 2.2~$\mu J/sample$ in onboard processing mode, thanks to an efficiency on the FFT computation task of 16.7~Mflops/s/mW with wireless bandwidth  reduction of 97\%, within a power budget of just 18.2~mW allowing for an operation time of 15~h.
\end{abstract}


\vspace{-0.1cm}
\section{Introduction}
\vspace{-0.1cm}

Recent wearable device design advancements are boosting the deployment of systems to monitor physiological parameters and support diagnostics for acute and chronic diseases. Wearables also represent a promising solution for unobtrusive human-machine interfaces by processing biosignals to enable bi-directional interaction between humans and machines \cite{ando2016wireless}.

Neural signals attract particular interest, given their potential for both clinical settings \cite{boonyakitanont2020review} and consumer-grade brain-computer interfaces (BCI) \cite{openbci}. 
Examples of wireless \ac{EEG} systems include the Emotiv Insight and EPOC+ \cite{emotiv}, Neurosky MindWave \cite{neurosky}, and OpenBCI \cite{openbci}. However, they lack on-board computational power and need to stream raw \ac{EEG} data (possibly filtered) to a benchtop computer for processing, thereby facing the limitation of the bandwidth of the low power data links \cite{BLE_throughput}. Only a few truly-wearable solutions are equipped with sufficient computation capabilities for onboard signal processing \cite{Kartsch2020,salvaro2018minimally}. However, they are typically operated for a single biopotential type and do not fuse information from heterogeneous biosensors.

Combining information from multiple biosignals demonstrated improved prediction capabilities for a variety of application scenarios. Examples include fusing \ac{EEG}, \ac{ECG}, and \ac{PPG} for drowsiness detection \cite{hong2018intelligent}; fusing \ac{EEG} and \ac{ECG} for epileptic seizures detection \cite{qaraqe2016epileptic}; fusing ultrasound (US) and \ac{EMG} for improved insights into muscle activity \cite{waasdorp2021combining}. These examples prove the need for wearable platforms capable of collecting and processing multiple heterogeneous biosignals in a wearable form factor. At the same time, the development of biosensor platforms with higher densities of biosignal inputs and more complex multi-sensor setups demands progressively higher edge-processing capabilities to keep the power budget acceptable. Following this development trend would allow to scale up the wearables, both in terms of single-device capabilities and in terms of multi-sensor body-networks.

This paper presents BioGAP, a \ac{PULP} platform for medical-grade acquisition and onboard processing of multiple heterogeneous biosignals. BioGAP is based on a modular design, with a baseboard (for managing power distribution, data flows, and processing) and expansion boards/modules for \ac{EEG} and \ac{PPG} data acquisition. The system comes in a small form factor (16$\times$21$\times$14 mm$^3$) and offers \ac{BLE} connectivity with a maximum throughput of 330 kbps. On-board processing capabilities are enabled by a \ac{PULP} \ac{SoC} from GreenWaves Technologies (GAP9), featuring ten RISC-V cores for state-of-the-art tiny-\ac{ML} inference~\cite{MLcommons}. The design modularity allows to further exploit BioGAP with a larger variety of biosensors (e.g., wearable ultrasound \cite{frey2022wulpus}).
The system is validated through \ac{PPG} measurements on the index finger and \ac{EEG} measurements of alpha waves and steady-state visual evoked potentials (SSVEP). Onboard data analysis is demonstrated using Fast Fourier Transform (FFT), executing eight \ac{FP} FFTs of size 1024 in less than 0.425~ms (102k cycles) and within a power budget of only 18.2 mW, which allows for 15~h of continuous operation with a 75~mAh battery.

\section{Description of the System}
\label{sec:methods}
\vspace{-0.1cm}

\subsection{Hardware: overview}
\begin{figure}[t]
\centerline{\includegraphics[width=0.88\columnwidth]{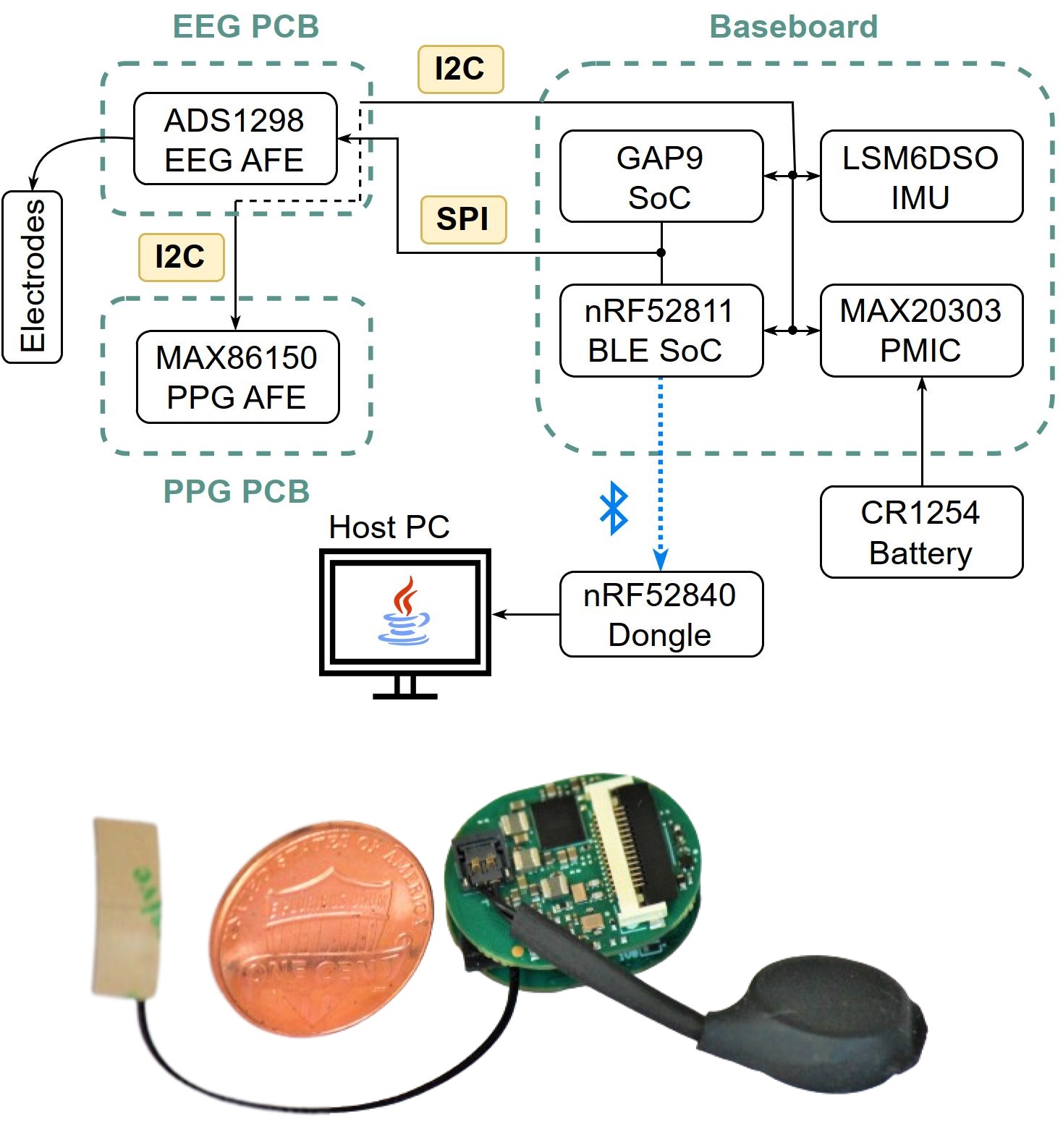}}
\vspace{-0.3cm}
\caption{System diagram (top) and photo (bottom) of the BioGAP platform next to a one-cent coin.}
\vspace{-0.4cm}
\label{fig:prototype}
\end{figure}

The design aimed for compact size, high onboard processing capabilities at a low power budget, modularity, and configurability. These features allow using BioGAP in widely different biopotential acquisition systems, ranging from the smallest size with relatively low channel count (e.g., earbuds) to high-density systems. BioGAP is based on (see Fig.~\ref{fig:prototype}):
\begin{itemize}
    \item \textbf{A baseboard} which controls the measurements, manages the power distribution, handles the data flow, and processing. It incorporates two SoCs: an nRF52811 (Nordic Semiconductor) for \ac{BLE} connection and a \ac{PULP} processor (GAP9, GreenWaves Technologies) for \ac{DSP} and \ac{NN} inferences at a low power envelope (up to 15.6 GOPs DSP and 32.2 GMACs machine learning at 370~MHz clock frequency). It also integrates volatile (256~Mbit) and non-volatile (128~Mbit) memories, an \ac{IMU} for movement sensing and device control, and a \ac{PMIC} (MAX20303, Analog Devices).
    \item \textbf{An 8-channel biopotential expansion board} (plugged as a shield on top of the baseboard) enabling EEG acquisition via one ADS1298 \ac{AFE} (Texas Instruments) that interfaces to the electrodes (both active and passive electrodes are supported).
    \item \textbf{A Pulse-Oximeter and \ac{HR} monitor module}, based on a MAX86150 (Maxim Integrated), connected to the EEG PCB via a \ac{FFC} to allow for flexibility in PPG sensor placement. 
\end{itemize}

\subsection{Hardware: baseboard}

\begin{figure}[tbp]
\centerline{\includegraphics[width=0.9\columnwidth]{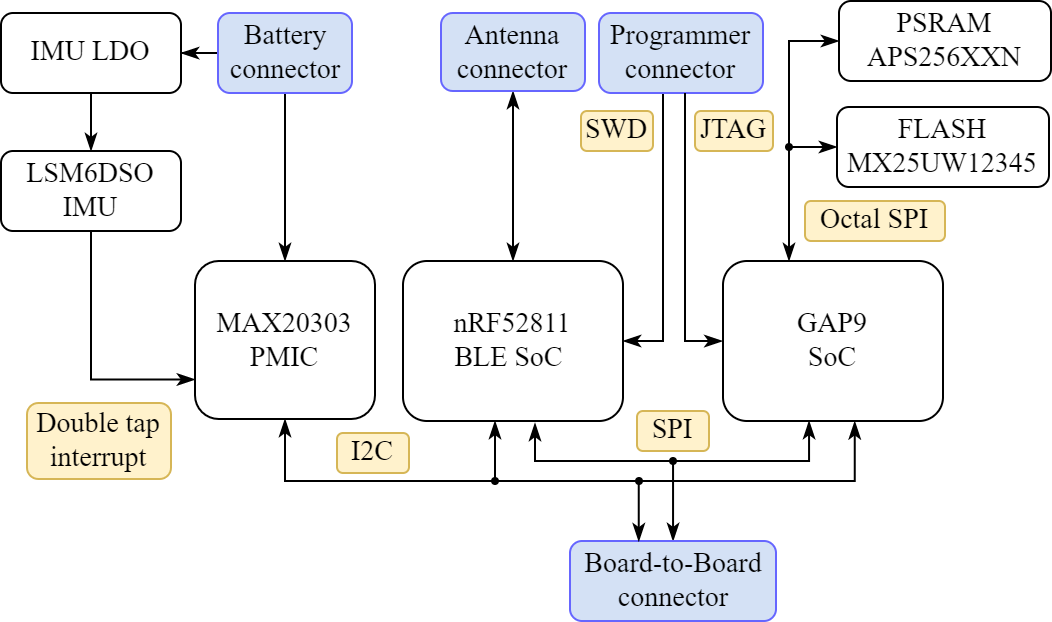}}
\vspace{-0.1cm}
\caption{Block diagram of the Baseboard.}
\vspace{-0.6cm}
\label{fig:BD_Baseboard}
\end{figure}

Figure~\ref{fig:BD_Baseboard} shows the baseboard block diagram.
The baseboard features an oval design and a size of only 16x21~mm$^2$.

\textbf{Interfaces, memories, sensors.}
Both SoCs have access to the main SPI and I2C bus to control the \ac{PMIC}, the \ac{IMU}, and the sensors (\ac{EEG}, \ac{PPG}). Only one SoC can be the bus master at a time, and a simple handshaking protocol is implemented between the two to allow them to request/surrender master privilege on the bus. The SPI bus is also used for data transfer between the two SoCs (e.g., to allow GAP9 to send/receive data via \ac{BLE}). Two high-speed octal SPI memories are available to GAP9 for data, code, and neural network weight storage. For volatile memory, we chose the APS256XXN-OBR PSRAM (AP Memory), which has a throughput of up to 400~MBps, a standby current of \SI{100}{\micro A} (typical), and offers different sleep modes with data retention. It was selected for its considerable capacity of 256~Mbit and compact size of 1.9x2.7~mm$^2$. 
For nonvolatile memory, the flash MX25UW12345G manufactured by Macronix was selected, offering 128~Mbit of storage space in a very compact package (3.5x3.8~mm$^2$). 
All the remaining interfaces of GAP9 (I2Cs, I3Cs, single/dual/quad SPIs, SAIs) except the MIPI-CSI2 camera interface are routed to the board-to-board connector to ensure the possibility of connecting the board to different types/numbers of sensors.
The baseboard also includes a low-power (\SI{26}{\micro A} accelerometer current consumption in LP mode with \SI{52}{Hz} ODR) IMU (LSM6DSO, ST Microelectronics) providing accelerometer, gyroscope, and temperature data. The IC has a small form factor (2.5x3~mm$^2$) and low power consumption (\SI{26}{\micro A} accelerometer current consumption in LP mode with \SI{52}{Hz} ODR). 
Furthermore, the IMU offers always-on low-power features such as tap recognition and free-fall detection, which are used for smart wake-up of BioGAP from deep-sleep and can be used to add additional smart features/control to the device.

\textbf{SoCs and radio.}
With data acquisition and processing from the sensors mostly handled by GAP9 SoC, the nRF52811 SoC from Nordic Semiconductors provides data communication capability implementing a flexible BLE communication at a low-power budget. The maximum power consumption of the radio is 13.8~mW in transmission and 15.6~mW when receiving data. Sleep states are available with 3.6 $\mu$W with full RAM retention and 1~$\mu$W with system OFF. The antenna is a 14x5.0x0.1~mm$^3$ FXP840 flexible antenna from Taoglas, a good fit for size-constrained applications.
GAP9 comes in a WLCSP 3.7x3.7~mm$^2$ package and combines a low power \ac{MCU}, a programmable compute cluster, and a hardware neural network accelerator, delivering up to 32.2 GMACs at 330 $\mu$W/GOP. It employs adjustable dynamic frequency and voltage scaling and automatic clock gating to tune the available compute resources and minimize power consumption for the given application. Deep sleep power consumption is 45 $\mu$W. GAP9 integrates the \ac{NE16}, which is a highly flexible hardware accelerator capable of handling various \ac{NN} operations, including \ac{CNN}, \ac{RNN}, and vector/matrix multiplications with 16- or 8-bit features and from 8 to 2 bit weights. It employs asymmetric, scaled quantization and the precision allocation can be customized for individual layers, allowing for fine-grained control to balance between network accuracy and energy efficiency in real-life applications. The architecture is equipped with a comprehensive kernel library optimized for neural networks and signal processing algorithms, making it readily applicable for various tasks. 

\begin{figure}[tbp]
\centerline{\includegraphics[width=0.75\columnwidth]{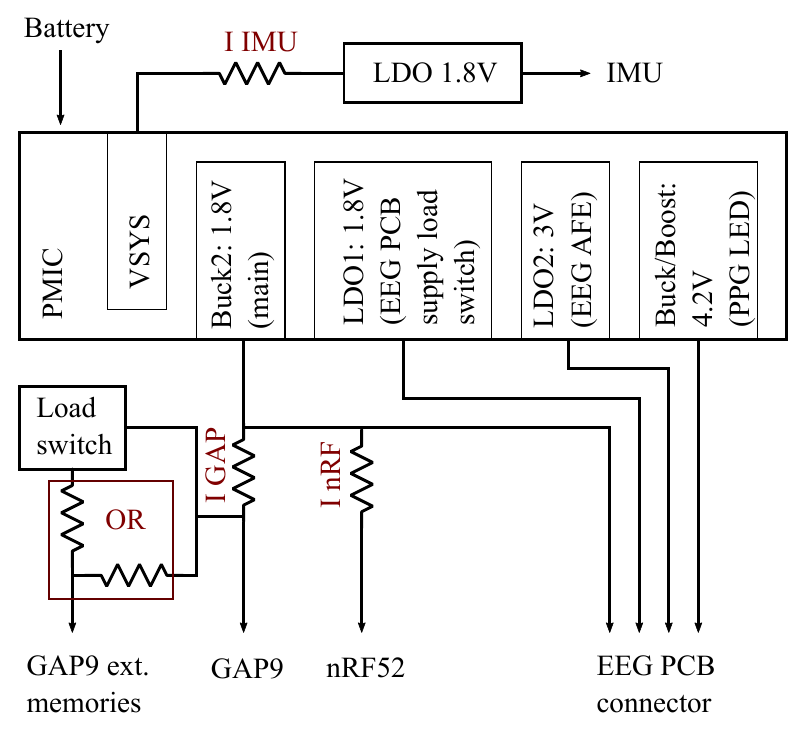}}
\vspace{-0.15cm}
\caption{Power domains implemented in the BioGAP platform. 
}
\vspace{-0.5cm}
\label{fig:BD_power_domains}
\end{figure}

\textbf{System power.}
The system is powered by a rechargeable coin battery (Varta CP1254 A4) weighing only 1.6~g and providing a capacity of 75~mAh. The different power domains and configurations are shown in Fig.~\ref{fig:BD_power_domains}. The SoCs are powered by the 1.8~V Buck2, a voltage available by default after the PMIC starts. To allow for more flexibility and energy efficiency, GAP9 memories can either be powered directly by the Buck2 or they can be powered through a load switch. The configuration can be changed with jumper resistors. The EEG and PPG PCBs main supply is provided by the PMIC integrated LDO1 (configured as a load switch) and can therefore be shut down completely when not needed. Furthermore, LDO2 is used to generate a low-noise analog supply for the EEG AFE. The LDO can be operated efficiently since the voltage difference between the battery and the 3~V analog supply is minimal. Finally, the Buck/Boost converter of the PMIC generates a 4.2~V supply for the PPG LEDs. The IMU is powered by a separate LDO to allow for an energy-efficient power save state, which is explained later in more detail. The individual power domains can be switched off to save energy when not required (e.g., it is possible to power down the analog part completely).

\subsection{Hardware: 8-channel biopotential expansion board}
\begin{figure}[tbp]
\centerline{\includegraphics[width=0.9\columnwidth]{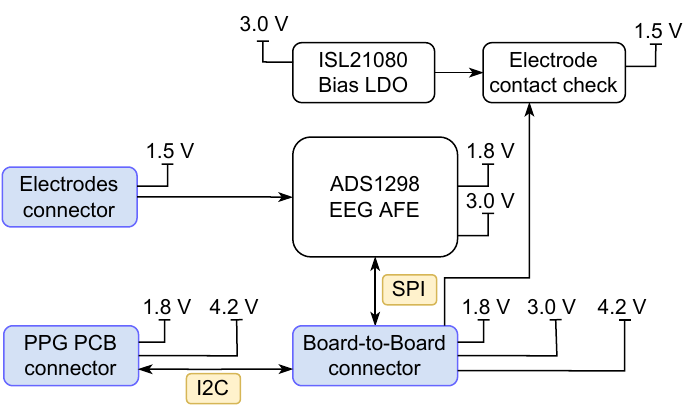}}
\vspace{-0.2cm}
\caption{Block diagram of the EEG PCB.}
\vspace{-0.5cm}
\label{fig:BD_EEG_PCB}
\end{figure}

Fig.~\ref{fig:BD_EEG_PCB} shows the block diagram of the complete board. The expansion board is connected to the Baseboard through a board-to-board connector that provides the needed voltages (1.8~V digital power, 3~V analog power, and 4.2~V for the PPG PCB).
The PCB allocates one ADS1298, a low-power 8-channel 24-bit AFE for bio-potential measurements. The AFE can be flexibly configured according to the application's needs, providing a programmable gain amplification (PGA) of 1 to 12 and data rates of 250 to 32~kSPS. Unused channels can be turned off. The EEG electrodes are interfaced through a compact DF52-12S-0.8H(21) board-to-wire connector, allowing the differential acquisition of up to 8 EEG channels with active or passive electrodes. 
The EEG PCB also measures electrode-skin contact impedance to check contact quality \cite{Kartsch2020}. 

\subsection{Hardware: Pulse-Oximeter and HR monitor
module}
Since the positioning of the PPG sensor is crucial for a good SNR, PPG measurement capabilities are implemented on a separate PCB, connected to the biopotential expansion board through an FFC cable to allow for the needed flexibility in sensor placement. The MAX86150 IC manufactured by Maxim was chosen for its compactness (3.3x5.6~mm$^2$) and the small amount of additional passive components needed to operate the IC. This allowed the design of a very compact PCB measuring only 9.2x7.2~mm$^2$. The MAX86150 is a fully integrated bio-signal sensor able to conduct PPG (pulse oximetry and heart rate sensor) measurements. It is operated with a 1.8~V supply (digital and analog) and a 4.2~V LED supply voltage and draws a current of \SI{480}{\micro A}~(typ.) at 10~SPS and \SI{1.15}{mA}~(typ.) at 100~SPS in SpO$_2$ mode.

\subsection{Operating principle and signal flow}
BioGAP can be used in the following modalities:

  \textbf{1) Streaming mode:} The acquired EEG data is continuously transmitted to a PC. It is then visualized in real-time, saved in a log file, and can be used for further processing. 
  The EEG and PPG data are read through DMA-enabled SPI (EEG data) and I2C (PPG data) interfaces. The user can choose different measurement configurations resulting in different buffering modalities or, in other words, different BLE packets. The buffer's content depends on the number of used EEG channels and the measurement mode (EEG, PPG, or EEG\&PPG). A ring buffer that holds up to 15 packets helps maintain wireless connectivity despite short-term interference or variable signal strength. \\
  
  \textbf{2) Computation on the edge mode:} The BioGAP directly acquires and processes EEG data, eliminating the need to transmit raw data wirelessly. 
  Instead, the bulk of the data is processed on the edge with the PULP chip and only a small amount of results is sent back to a dongle and from there to a PC. In this mode, the GAP9 takes over the control of the SPI bus as a master and acquires the sensor data directly.\\
  
  \textbf{3) Sleep mode:} The PMIC turns off all its main regulators and powers down the two SoCs to minimize power consumption ($<150~\mu W$, $>70~days$ battery lifetime). The IMU is the only device powered by the PMIC to allow the detection of double-tap events. Upon its detection, an interrupt from the IMU signals the PMIC to turn on its Buck2 converter and therefore the SoCs, exiting the power save state.\\
  
\textit{Operation:} after starting up, BioGAP establishes a BLE connection with the receiver Dongle, goes into a low-power BLE connected state, and waits for further commands. 
A Java GUI is used to display measurements in real-time and to send commands (switch between streaming, computation on the edge, and sleep mode; start or stop measurement) and parameters (number of active EEG and PPG channels; sampling frequency; EEG gain) to the BioGAP through the Dongle and BLE.
The BioGAP uses the BLE connection to stream out the raw data of measurements (streaming mode) or the results of onboard computations (computation on the edge mode).
When the respective command is received, BioGAP starts the biosignal measurements in one of the two mentioned modalities with the set parameters. Once the measurement is completed, the device returns to the BLE-connected state. When not used for a while, the platform can be sent into sleep mode through a BLE command, allowing it to save power. A double-tap gesture wakes up the system. 
\section{Experimental Verification}
\label{sec:results}
\vspace{-0.1cm}

\subsection{EEG acquisition}
We characterized the system operating the ADS1298 at 1~kSPS and a gain of 6 in high-resolution mode, with a \mbox{-3~dB} cutoff frequency of 262~Hz. The integrated RMS noise from 0.5 to 100~Hz is \SI{0.47}{\micro V}, which is consistent with IFCN standards for clinical recording of \ac{EEG} signals~\cite{nuwer1998ifcn}.
We validated BioGap AFE on a single subject through two traditional \ac{EEG} paradigms: alpha waves and steady-state visual evoked potentials (SSVEP). The \ac{EEG} setup includes a single channel placed at POz, with GND and Reference at A1 and A2 (10-20 reference system). Tests were performed in a quiet environment with dimmed light and away from well-known sources of electrical interference. The subject reported no neurological or psychiatric disorders and gave informed consent for inclusion before participating in the study.

\subsubsection{Alpha waves} 

\begin{figure}[t]

\centerline{\includegraphics[width=0.95\columnwidth]{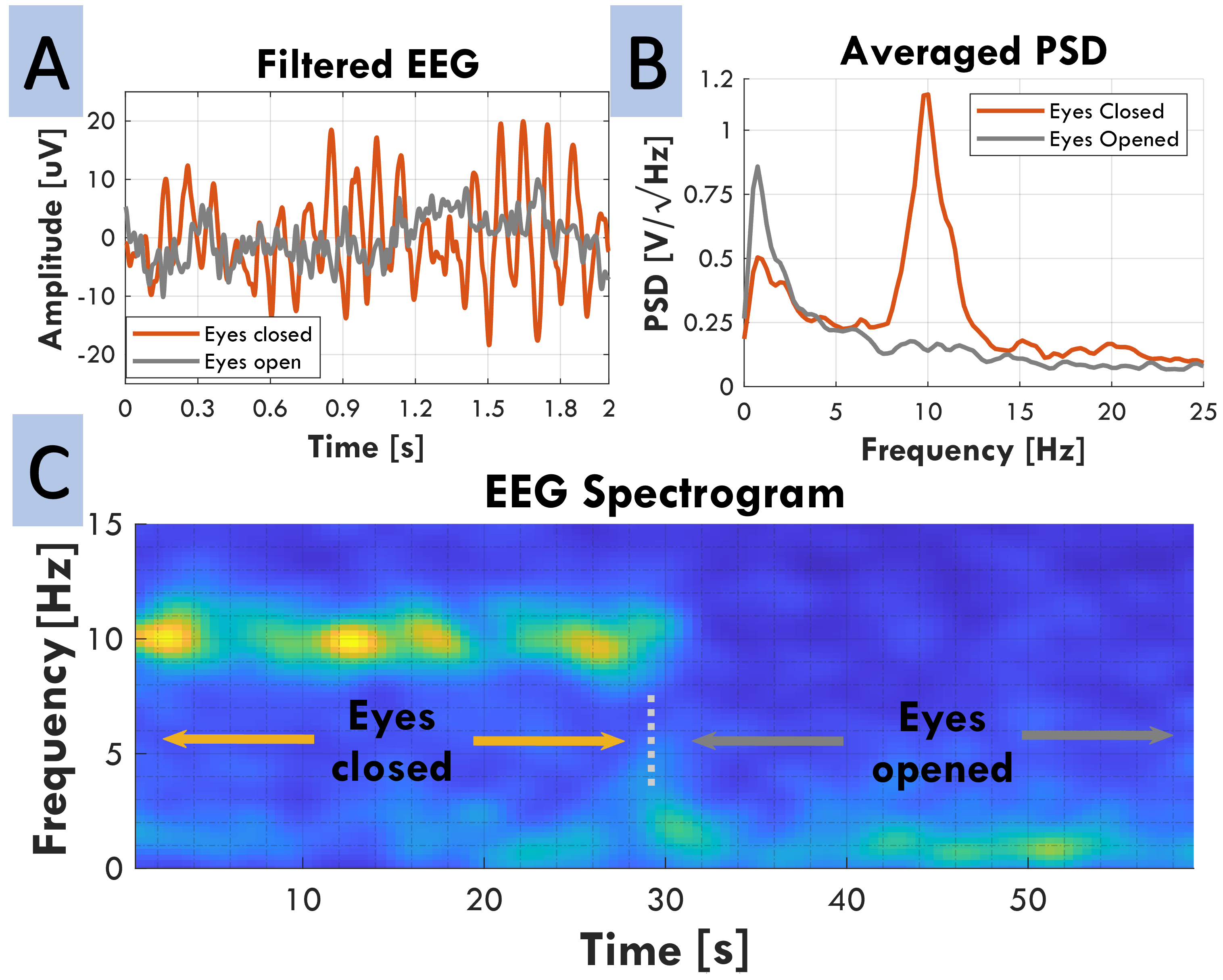}}
\vspace{-0.3cm}
\caption{Measurement of the alpha wave activity in eyes closed vs eyes open condition.}
\vspace{-0.6cm}
\label{fig:alpha_waves}
\end{figure}

Besides medical and HMI applications, alpha waves, i.e., their presence/absence during open/close eye conditions, are typically used as the first qualitative marker for \ac{EEG}-based systems. Specifically, during the test, the subject alternates between closed/open eyes state (30s per state) to elicit alpha waves \cite{bazanova2014interpreting}. 
Fig. \ref{fig:alpha_waves}a shows the \ac{EEG} response for open/close conditions in the time domain, indicating significant amplitude differences between the open and close eyes conditions (approx. 13 and 37 uVpp, respectively). Fig. \ref{fig:alpha_waves}b shows the corresponding Power Spectral Density (PSD) over 30~s for both conditions, showing a 4x power increase for the closed-eyes state. Fig. \ref{fig:alpha_waves}c shows the signal spectrogram (1024-sample windows with a 768-sample overlap) computed over a signal segment where a closed/open eye transition occurs, illustrating a high energy component for the alpha band and a subsequent decrease when the subject opens its eyes.

\subsubsection{SSVEP}\label{subsec:ssvep}
\begin{figure}[bp]
\vspace{-0.4cm}

\centerline{\includegraphics[width=0.9\columnwidth]{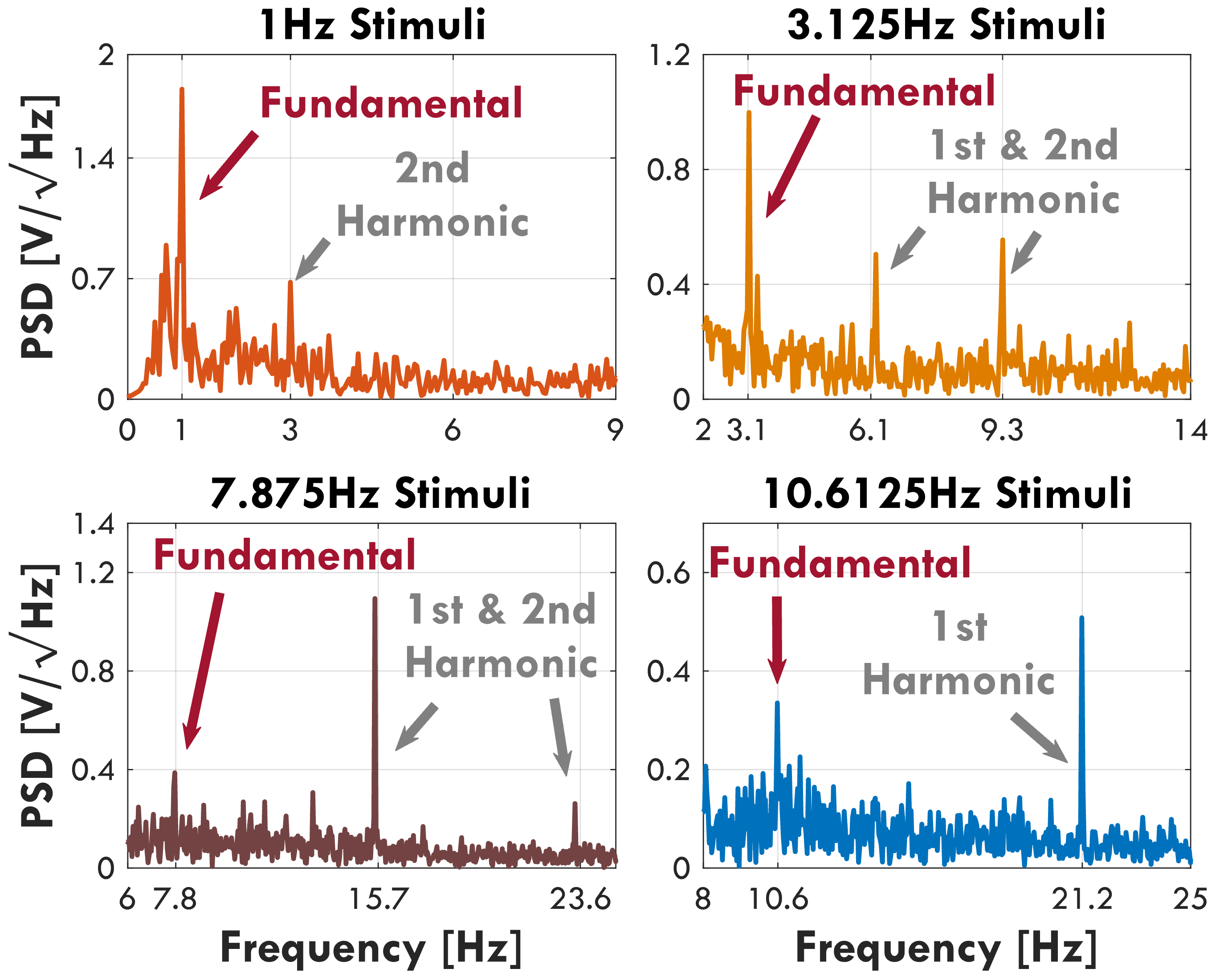}}
\vspace{-0.2cm}
\caption{Spectrum of an \ac{EEG} measurement during SSVEP stimulation with different frequencies.}
\label{fig:ssvep_figure}
\end{figure}
Several SoA BCI systems rely on visual stimuli, such as steady-state visual evoked potentials, whose processing by the visual cortex is typically detected through frequency-domain analysis of the elicit \ac{EEG}. In light of this, we tested our system through a series of consecutive stimuli consisting of a sinusoidal on-off 100\% contrast temporal modulation black and white \mbox{10~{$\times$} 10} square checkerboards. Stimuli include four frequencies (1, 3.125, 7.8125, and 10.6125~Hz) presented in a randomized order three times. Each trial lasts 25~s and is always followed by a 10~s rest period to avoid visual fatigue. 

Fig.~\ref{fig:ssvep_figure} presents the Power Spectral Density of averaged (per stimulation frequency) trials. In all cases, SSVEP response is present, including 1~Hz, the most challenging case, typically masked by intrinsic brain activity and noise. The remaining frequencies not only show a good response to the stimuli frequency but also to signal harmonics, which can be further exploited to better identify a given frequency in a BCI context.

\subsection{PPG acquisition}
\begin{figure}[tbp]
\centerline{\includegraphics[width=0.9\columnwidth]{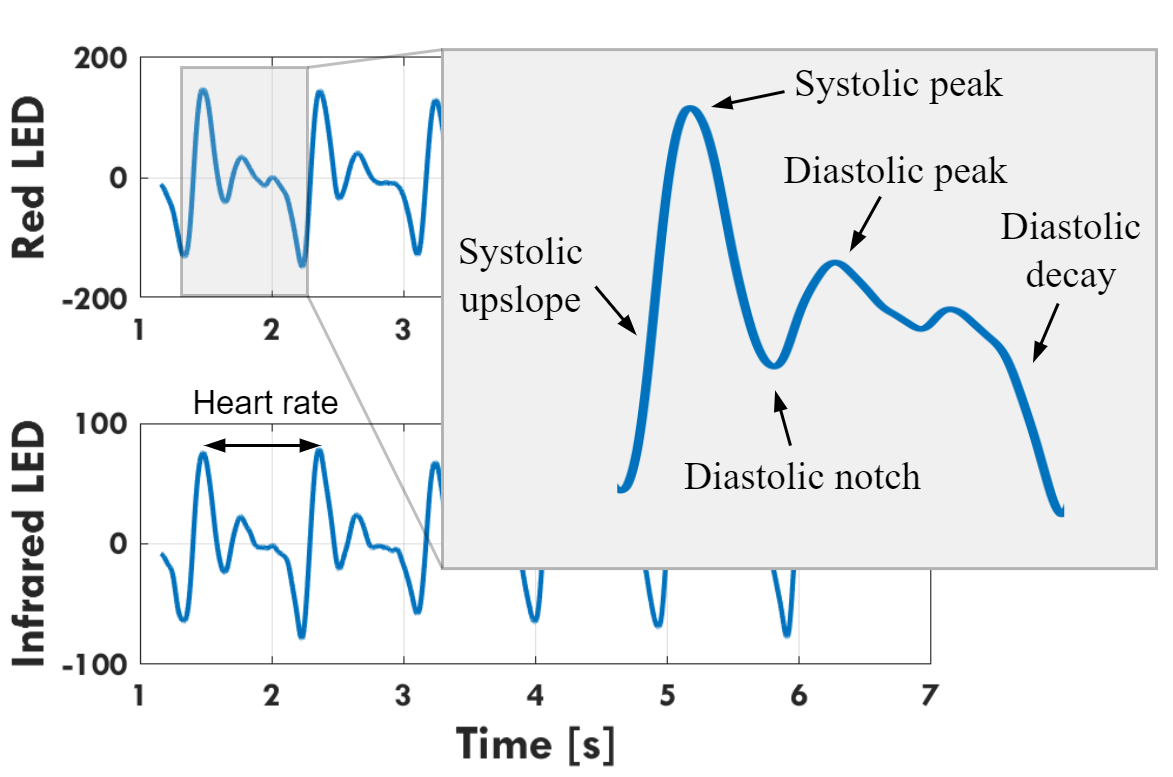}}
\vspace{-0.3cm}
\caption{\ac{PPG} signal measured at the index finger with the red (top) and IR (bottom) LED. %The different phases of the cardiac cycle can be easily identified.}
}
\vspace{-0.3cm}
\label{fig:ppg_meas}
\end{figure}

Fig.~\ref{fig:ppg_meas} shows a typical \ac{PPG} measurement collected at the index finger via a red LED (top) or infrared (IR) LED (bottom). The inset highlights the phases of the cardiac cycle. After subtracting the moving mean, the raw signal was filtered with a Gaussian averaging filter with a window size of 0.1~s.

\subsection{Power measurements}
BioGAP operates in different modes to optimally balance power and performance. Individual power domains are characterized by measuring the voltage drop over 1 Ohm current shunt resistors using a Keysight 34465A digital multimeter. 
Unless otherwise stated, the ADS1298 is operated in high-resolution mode at 1~kSPS with a PGA gain of 6 and 8 active channels for \ac{EEG} acquisition. 
\subsubsection{Streaming mode}
\begin{figure}[tbp]
\centerline{\includegraphics[width=0.9\columnwidth]{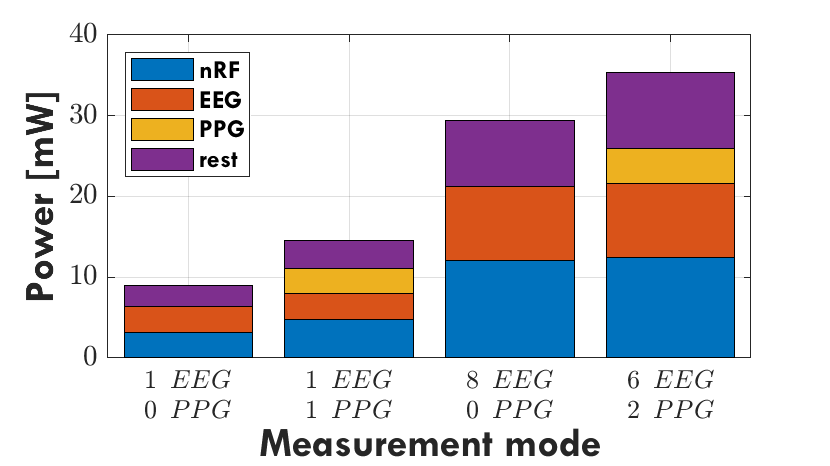}}
\vspace{-0.2cm}
\caption{Power measurement of BioGAP in streaming mode for different EEG \& PPG configurations. Sampling at 1 kSPS.
}
\vspace{-0.5cm}
\label{fig:meas_power_streaming_mode}
\end{figure}

Fig.~\ref{fig:meas_power_streaming_mode} reports the power consumption of the individual power domains in the streaming mode with different numbers of active \ac{EEG} and \ac{PPG} channels. The MAX86150 operates at 100~SPS. In all cases, the consumption is dominated by the 3.0~V analog domain which supplies the ADS1298 AFE and the nRF52 SoC (mainly for streaming the data through BLE). The power consumption of the ADS1298 is inherently linked to the nature of the \ac{EEG} signal demanding a high-resolution acquisition at very low noise and hundreds of Hz of sampling frequency.

\begin{table*}[ht]
\begin{center}
\begin{threeparttable}[b]
\caption{Performance comparison to state-of-the-art wearable devices for biosignal processing.}
\label{table:competitors)}
\scriptsize{
\begin{tabular}
{
p{0.5in}>
{\centering\arraybackslash}p{1.1in}>
{\centering\arraybackslash}p{0.7in}>
{\centering\arraybackslash}p{0.4in}>
{\centering\arraybackslash}p{0.6in}>
{\centering\arraybackslash}p{0.9in}>
{\centering\arraybackslash}p{0.7in}>
{\centering\arraybackslash}p{0.8in}>
{\centering\arraybackslash}p{1.0in}
}

\toprule[0.20em]

\textbf{Device} & \textbf{~~~~~Biosignal  \newline (Application)} & \textbf{~~~Size \newline [mm x mm]} & \textbf{\# ch.} & \textbf{Connectivity} &\textbf{SoC}  &\textbf{Performance [Mop/s]} & \textbf{Energy efficiency ~[Mop/s/mW]}\\
\midrule
(int32) 

\cite{kaveh2020wireless} & EEG (EOG, alpha, ASSR) & 25 $\times$ 25 & 5$^1$ & BLE & ARM (SmartFusion2) & 166 (int32) & n/a%2.17 
\\
\midrule[0.12em]
\cite{salvaro2017towards} & EEG (SSVEP) & 45 $\times$ 90 & 3$^2$ & BT2.1  & ARM-STM32F407 & 168 (FP32) & 1.2 (FP32) %6.10 \\
\\
\midrule[0.12em]

\cite{Wang2021_Sleep_EEG} & EEG (SpEn) & 34 $\times$ 27 & 4 & BLE & Nordic nRF52840 & 64 (FP32) & 6.4 (FP32) %37 \\
\\
\midrule[0.12em]
\cite{guermandi2022wireless} & EEG (AEP) & 15 $\times$ 16 & 1$^3$ & BLE & Nordic nRF52811 & 64 (int32) & 9.7 (int32) %16 \\
\\
\midrule[0.12em]

\cite{Kartsch2020} & EEG (SSVEP) & 20 $\times$ 40 & 3$^2$ & BLE & Mr.Wolf & 1'000 (FP32)$^5$ & 18 (FP32) %18 \\
\\
\midrule[0.12em]

\textbf{BioGAP} \newline \textbf{(this work)} & \textbf{~~~~EEG (SSVEP)} \newline \textbf{PPG} & \textbf{16 $\times$ 21} & \textbf{8 EEG} \newline \textbf{2 PPG$^4$} & \textbf{BLE}  & \textbf{GAP9} & \textbf{3'000 (FP32)$^{5, 6}$} & \textbf{74 (FP32)} \\
\midrule[0.12em]
%\bottomrule
 
\end{tabular}}
AEP: auditory evoked potentials, ASSR: Auditory Steady-State Response, SpEn: Sample Entropy, BT: Bluetooth, EOG: electrooculography, FP: floating point \\ (1) up to 64 possible, (2) up to 8 possible, (3) up to 3 possible, (4) time-multiplexed, (5) 2 FLOPSs = 1 F32MAC, (6) at 370~MHz. \\

\end{threeparttable}
\end{center}
\vspace{-6mm}
\end{table*}

\subsubsection{Computation on the edge mode}
The nRF52 SoC's primary contribution to power consumption can be significantly reduced by conducting most of the computation on the device and only transmitting a few results. To demonstrate this, we employed the classical SSVEP \ac{EEG} setting described in section~\ref{subsec:ssvep} with FFT analysis as it is one of the most used DSP algorithms and therefore serves as a baseline for comparison. The SSVEP response can be obtained by analyzing the relevant frequency bins (fundamental frequency and the first two harmonics for each of the four stimulation frequencies of the SSVEP experiment) of the spectrum. The power in the relevant frequency bins is obtained by computing real FFTs of the time signal (one FP FFT per channel every 50~ms resulting in a suitable latency for \ac{BCI} applications) and transmitted to the computer via the nRF52 and BLE. Setting the GAP9 core voltage to 0.65~V and the compute cluster frequency to 240~MHz resulted in optimal power efficiency. The FFTs are computed one after another in parallel on eight cluster cores achieving a task energy efficiency of 16.7~Mflops/s/mW. One \ac{FP} FFT of size 1024 requires less than 13k cycles for completion, achieving a parallel speedup of 5.3$\times$. The computation of the eight FFTs takes 0.425~ms, including the DMA transfers that prepare the data for the subsequent FFT. This results in an energy efficiency per sample of more than 2.2~$\mu J$, compared to 3.6~$\mu J$ in the streaming mode.

\subsubsection{Comparison streaming and computation on the edge}
\begin{figure}[tbp]
\vspace{-0.4cm}
\centerline{\includegraphics[width=0.9\columnwidth]{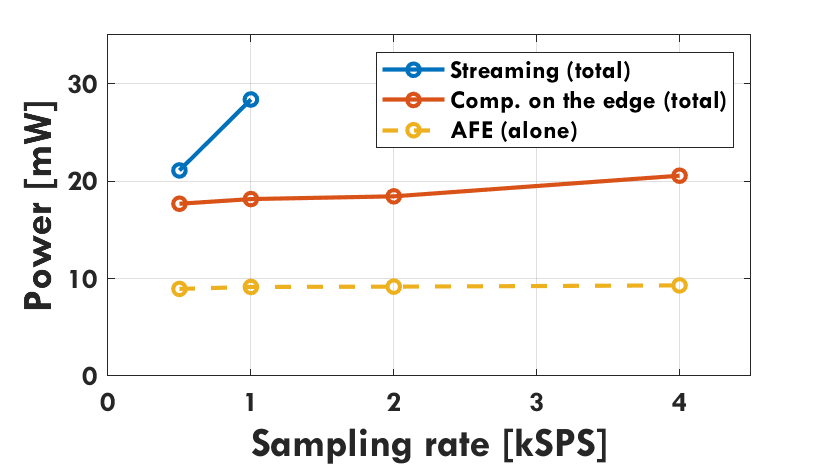}}
\vspace{-0.2cm}
\caption{Power measurement of BioGAP in streaming and computation on the edge mode at different sampling rates. The contribution of the AFE is represented by the yellow line and is the same for both modes.}
\vspace{-0.5cm}
\label{fig:meas_power_GAP_vs_BLE}
\end{figure}

We compare streaming and computation on the edge mode by varying the \ac{EEG} sampling frequency for both modes. 
To cover the same time window of 1~s at the various sampling frequencies, the FFT size is adjusted accordingly (e.g., a 4~kSPS sampling rate corresponds to an FFT size of 4096, while 2~kSPS corresponds to size 2048). The FFT overlap is chosen such that a new set of FFT results (one FFT per channel, eight in total) is computed every 50~ms.
Fig.~\ref{fig:meas_power_GAP_vs_BLE} shows the power consumption in the two modalities at different frame rates. 
The power numbers in the streaming mode stop at 1~kSPS, due to BLE limitations (the effective payload throughput is currently limited to 330~kbps for unsupported data length extension and packet headers overheads), preventing the use of the platform in this mode at higher sampling frequencies.
Hence, scaling up the system (e.g. using a higher sampling frequency or adding more sensors) is solely possible through energy-efficient onboard processing. 
Furthermore, even at lower \ac{BW} such as 192~kbps, corresponding to the application level throughput needed to stream eight channels of EEG data at 1~kSPS, it can be difficult to reliably sustain an uninterrupted BLE connection due to short-time interference from other devices. This underlines another advantage of processing sensor data on the edge, as this reduces the needed application level \ac{BW} to 5.12~kbps (1 \ac{FP} result per channel every 50~ms) which translates to a 97~\% reduction in the needed \ac{BW}.

\subsection{Comparison to related works}

Table \ref{table:competitors)} summarizes the positioning of BioGAP to related works for biosignal processing. BioGAP offers the most significant flexibility for signal acquisition within the compared devices, allowing using the same device for different applications or sensor fusion. BioGAP also achieves the smallest device size within our comparison, allowing for non-stigmatized usage. Finally, as regards onboard computation capabilities, BioGAP allows 2$\times$ higher performance with respect to the closest competitor \cite{Kartsch2020} at 4$\times$ higher energy efficiency. It is also important to note that \cite{Kartsch2020} is based on a research prototype SoC, while BioGAP is based on components available in volume production.

\section{Conclusion}
\vspace{-0.1cm}

In this paper, we presented BioGAP, a compact (16x21x14~mm$^3$, 6g) and versatile biosignal acquisition platform. We demonstrated its \ac{EEG} functionality by means of two traditional \ac{EEG} paradigms: alpha waves and SSVEP (both in streaming and computation on the edge mode). The \ac{PPG} capabilities are demonstrated with a measurement at the index finger. 
The presented platform integrates a SoA PULP SoC (GAP9) featuring ten RISC-V cores, which enabled onboard computation of eight 1024 floating point FFTs requiring only 0.425 ms and continuous operation in on-the-edge computation mode of 15 h.
BioGAP effectively brings increased SoA computation power on the edge, and can serve as a base for the development of novel biosensor platforms with higher energy-efficient on-board intelligence.

Future work will focus on employing BioGAP for heterogeneous wearable biosignal applications. The use of GAP9 will allow for energy-efficient machine learning inference at the edge for small form factor wearables, opening a range of new possibilities from sensor fusion approaches (e.g. \ac{EEG}, \ac{PPG}, and ultrasound) to high density \ac{EEG} applications. The platform's modularity also allows adding new components (e.g., for a higher \ac{EEG} channel count) or adjusting existing ones (e.g., to tailor the physical dimensions to a specific application such as in-ear \ac{EEG} and \ac{PPG}).

\section*{Acknowledgment}
\vspace{-0.1cm}
We thank A. Blanco Fontao and H. Gisler (ETH Zürich) for technical support.

\bibliographystyle{IEEEtran}
\bibliography{bibliography.bib}
\end{document}